\pdfoutput=1
\PassOptionsToPackage{table}{xcolor}

\documentclass[sigplan,screen]{acmart}

\usepackage{xspace}
\usepackage{enumitem}
\usepackage{booktabs, multirow} %
\usepackage{soul}%
\usepackage[ruled, vlined, linesnumbered, commentsnumbered, longend]{algorithm2e}
\usepackage{graphicx}
\usepackage{subcaption} %
\usepackage[normalem]{ulem}
\usepackage{tcolorbox}
\usepackage[table]{xcolor}
\usepackage{listings}
\usepackage{dutchcal}
\usepackage{balance} 

\usepackage{tikz}
\newcommand*\circled[1]{\tikz[baseline=(char.base)]{
            \node[shape=circle,fill,inner sep=0.5pt] (char) {\textcolor{white}{#1}};}}
\definecolor{applegreen}{rgb}{0.55, 0.71, 0.0}

\definecolor{darkgreen}{rgb}{0.0, 0.5, 0.0}

\copyrightyear{2025}
\acmYear{2025}
\setcopyright{cc}
\setcctype{by}
\acmConference[ASPLOS '25]{Proceedings of the 30th ACM International Conference on Architectural Support for Programming Languages and Operating Systems, Volume 2}{March 30-April 3, 2025}{Rotterdam, Netherlands}
\acmBooktitle{Proceedings of the 30th ACM International Conference on Architectural Support for Programming Languages and Operating Systems, Volume 2 (ASPLOS '25), March 30-April 3, 2025, Rotterdam, Netherlands}\acmDOI{10.1145/3676641.3716022}
\acmISBN{979-8-4007-1079-7/25/03}

\newcommand{\revision}[1]{#1}

\newcommand{\CodeIn}[1]{{\small\texttt{#1}}}
\newcommand{\Comment}[1]{}
\newcommand{\parabf}[1]{\noindent\textbf{#1}}

\newcommand{\eg}{\emph{e.g.,}\xspace}
\newcommand{\ie}{\emph{i.e.,}\xspace}

\newcommand{\llm}{{LLM}\xspace} 
\newcommand{\gpt}{{GPT}\xspace}

\newcommand{\syzkaller}{{Syzkaller}\xspace} 
\newcommand{\syzlang}{{syzlang}\xspace} 

\newcommand{\difuse}{{DIFUSE}\xspace} 
\newcommand{\syzdescribe}{{SyzDescribe}\xspace}

\newcommand{\titanfuzz}{{TitanFuzz}\xspace} 
 
\newcommand{\fuzzall}{{Fuzz4All}\xspace}

\newcommand{\llvm}{{LLVM}\xspace} 
\newcommand{\kernel}{{Linux kernel}\xspace} 

\newcommand{\tech}{{KernelGPT}\xspace}

\newcommand{\totalDriver}{666\xspace}
\newcommand{\indevDriver}{278\xspace}
\newcommand{\missingDriver}{75\xspace}
\newcommand{\validDriverNoRep}{40\xspace}
\newcommand{\repairDriver}{30\xspace}
\newcommand{\validDriver}{70\xspace}

\newcommand{\totalSocket}{85\xspace}
\newcommand{\indevSocket}{81\xspace}
\newcommand{\missingSocket}{66\xspace}
\newcommand{\validSocketNoRep}{45\xspace}
\newcommand{\repairSocket}{12\xspace}
\newcommand{\validSocket}{57\xspace}

\newcommand{\numNewSpec}{532\xspace}
\newcommand{\newPercent}{13.6\%\xspace}
\newcommand{\numNewType}{294\xspace}
\newcommand{\numSyzSpec}{3903\xspace}

\newcommand{\numNewBug}{24\xspace}
\newcommand{\numCVE}{11\xspace}
\newcommand{\numConfirmed}{21\xspace}
\newcommand{\numFix}{12\xspace}

\newcommand{\numBugNewDriver}{17\xspace}
\newcommand{\numBugNewSpec}{7\xspace}

\newcommand{\numDescNewSpec}{146\xspace}
\newcommand{\numDescNewPercent}{3.7\%\xspace}
\newcommand{\numDescNewType}{168\xspace}

\newcommand{\moreSocketLine}{18.6}

\newcommand{\Challenge}[1]{L-{#1}\xspace}
\newcommand{\Solution}[1]{Solution\xspace}

\begin{document}

\title{\tech: Enhanced Kernel Fuzzing via Large Language Models}

\author{Chenyuan Yang}
\orcid{0000-0002-7976-5086}
\affiliation{%
  \institution{University of Illinois at Urbana-Champaign}
  \city{Champaign}
  \country{USA}
}
\email{cy54@illinois.edu}

\author{Zijie Zhao}
\orcid{0009-0008-0718-3088}
\affiliation{%
  \institution{University of Illinois at Urbana-Champaign}
  \city{Champaign}
  \country{USA}
}
\email{zijie4@illinois.edu}

\author{Lingming Zhang}
\orcid{0000-0001-5175-2702}
\affiliation{%
  \institution{University of Illinois at Urbana-Champaign}
  \city{Champaign}
  \country{USA}
}
\email{lingming@illinois.edu}



\begin{abstract}
Bugs in operating system kernels can affect billions of devices and users all over the world. As a result, a large body of research has been focused on kernel fuzzing, i.e., automatically generating syscall (system call) sequences to detect potential kernel bugs or vulnerabilities.
Kernel fuzzing aims to generate valid syscall sequences guided by syscall specifications that define both the syntax and semantics of syscalls.
While there has been existing work trying to automate syscall specification generation, this remains largely manual work, and a large number of important syscalls are still uncovered.

In this paper, we propose \tech, the first approach to automatically synthesizing syscall specifications via Large Language Models (LLMs) for enhanced kernel fuzzing.
Our key insight is that LLMs have seen massive kernel code, documentation, and use cases during pre-training, and thus can automatically distill the necessary information for making valid syscalls.
More specifically, \tech leverages an iterative approach to automatically infer the specifications, and further debug and repair them based on the validation feedback.
Our results demonstrate that \tech can generate more new and valid specifications and achieve higher coverage than state-of-the-art techniques.
So far, by using newly generated specifications, \tech has already detected \numNewBug new unique bugs in Linux kernel, with \numFix fixed and \numCVE assigned with CVE numbers.
Moreover, a number of specifications generated by \tech have already been merged into the kernel fuzzer \syzkaller, following the request from its development team.
\end{abstract}

\settopmatter{printacmref=true}
\maketitle

\section{Introduction}

Operating system kernels are among the most critical systems, as all other types of systems rely on and operate on them.
Kernel vulnerabilities, such as crashes or out-of-bounds writes, can be maliciously exploited, potentially causing substantial harm to all users.
To ensure the correctness and security of these fundamental systems, fuzzing (or fuzz testing)~\cite{sutton2007fuzzing,zeller2019fuzzing,yang2023fuzzing} has been employed for decades.
Such techniques automatically generate a vast number of system calls as test inputs, intending to detect potential kernel bugs.

Among various kernel fuzzing techniques~\cite{kernel-fuzzing,Triforce,trinity}, \syzkaller~\cite{syzkaller} is one of the most popular tools. \syzkaller has identified over 5K bugs that are recognized and fixed by kernel developers~\cite{syzbot}.
Thus, numerous research efforts have focused on enhancing \syzkaller, targeting areas such as seed generation~\cite{moonshine,enriched-corpus}, seed selection~\cite{syzvegas}, guided mutation~\cite{healer,fleischer2023actor}, and syscall specification generation~\cite{difuze,syzdescribe,syzgen,ksg}.
Among these, the syscall specifications written in \syzlang~\cite{syzlang} are particularly crucial, significantly contributing to the effectiveness of \syzkaller and allowing it to cover more kernel modules.
They specify the syntax of syscalls, and their intra- and inter-dependencies, enabling the generation of more valid syscall sequences that probe deeper into the kernel code logic.
However, crafting syscall specifications is difficult because it is predominantly a manual process and require much in-depth kernel knowledge.

\begin{figure*}
    \centering
    \includegraphics[width=0.9\linewidth]{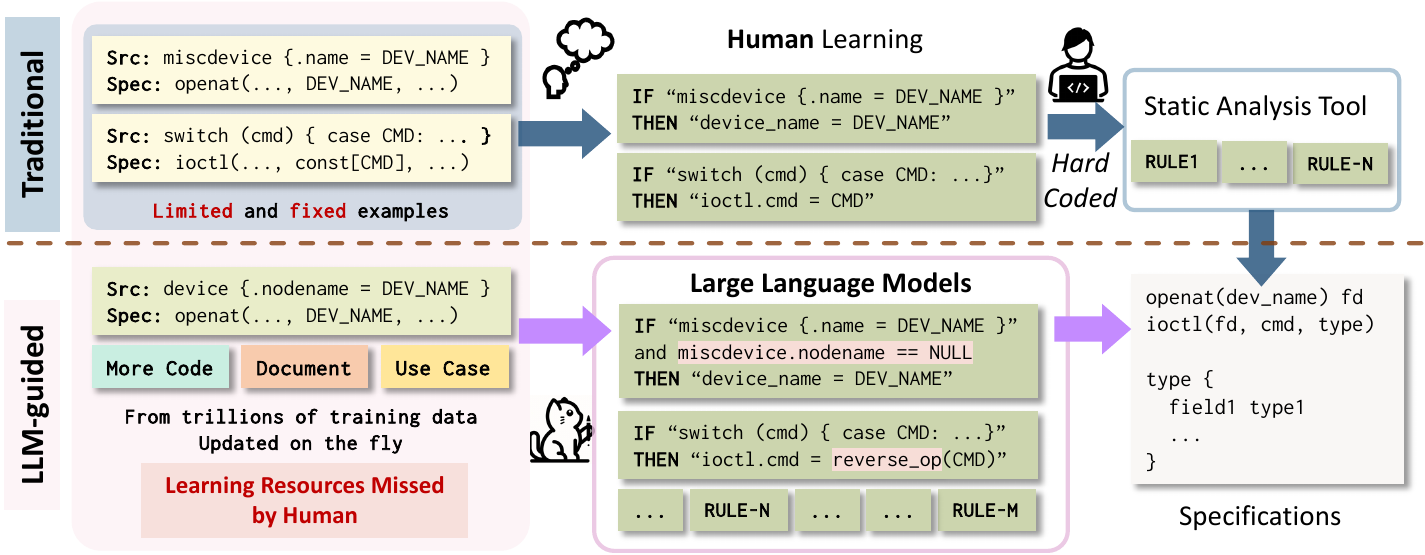}
    \caption{Workflows of syscall specification inference based on static analysis and \llm{s}}
    \label{fig:intro}
\end{figure*}

To address this issue, recent research has focused on automating the generation of syscall specifications, particularly for device drivers.
For instance, \difuse~\cite{difuze} and \syzdescribe~\cite{syzdescribe} employ static code analysis to identify device driver syscall handlers and infer their corresponding descriptions.
The top half of Figure~\ref{fig:intro} illustrates the workflow of static analysis-based techniques.
Initially, experts manually define rules to infer descriptions from the source code, drawing upon their own understanding of the kernel codebase and existing \syzkaller examples.
The accuracy and effectiveness of the generated syscall descriptions depend heavily on the comprehensiveness of these mapping rules, which is often challenging, costly, and tedious.
Moreover, as the kernel codebase evolves, these mapping rules are subject to frequent changes.
Keeping up with these evolving scenarios is a significant challenge for static analysis methods, particularly given the extensive implementation efforts involved.
Plus, existing approaches struggle to generate human-readable specifications, yet readability is essential for validation and maintenance, according to \syzkaller developers \cite{syzkaller-spec-gen-issue}.

\begin{figure}[h]
    \centering
    \captionsetup[subfigure]{justification=centering, aboveskip=3pt, belowskip=2pt}
    \begin{subfigure}{\linewidth}
        \centering
        \includegraphics[width=0.85\linewidth]{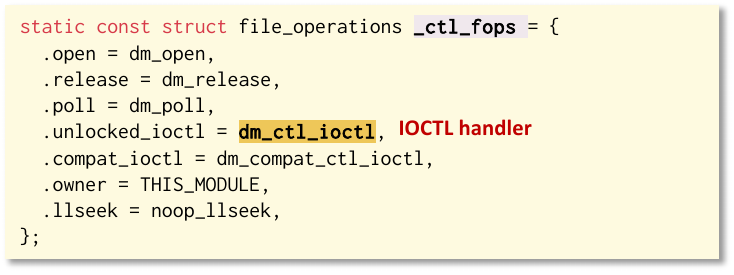}
        \caption{Device operation handler}
        \label{fig:dm-ctl-example-a}
    \end{subfigure}

    \begin{subfigure}{\linewidth}
        \centering
        \includegraphics[width=0.85\linewidth]{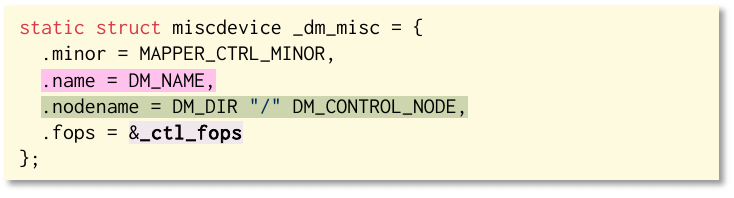}
        \caption{Device operation handler reference}
        \label{fig:dm-ctl-example-b}
    \end{subfigure}

    \begin{subfigure}{\linewidth}
        \centering
        \includegraphics[width=0.85\linewidth]{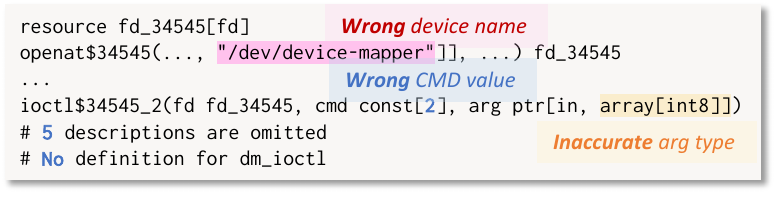}
        \caption{Specification generated by \syzdescribe}
        \label{fig:dm-ctl-example-c}
    \end{subfigure}

    \begin{subfigure}{\linewidth}
        \centering
        \includegraphics[width=0.85\linewidth]{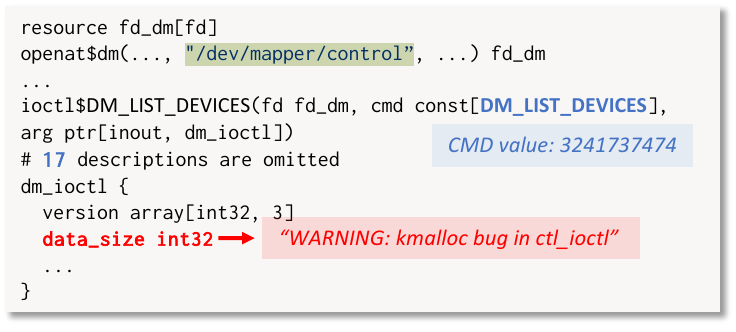}
        \caption{Specification generated by \tech}
        \label{fig:dm-ctl-example-d}
    \end{subfigure}
    \captionsetup{aboveskip=5pt}
    \caption{Device mapper driver in \texttt{drivers/md/dm-ioctl.c}}
    \label{fig:dm-ctl-example}
\end{figure}

Take, for instance, Figure~\ref{fig:dm-ctl-example-a} and Figure~\ref{fig:dm-ctl-example-b}, which illustrates the source code of two \CodeIn{struct} variables, associated with the device mapper driver~\cite{device-mapper}, responsible for mapping physical block devices to higher-level virtual block devices.
Specifically, these two variables are the device operation handler and its reference usage, crucial for inferring the device name.
Current advanced syscall description generators, like \syzdescribe~\cite{syzdescribe}, typically rely on the field \CodeIn{name} in \CodeIn{struct miscdevice} to determine the device name for driver interaction, which is a conventional use case. %
However, in this example, the correct device name is actually specified in the field \CodeIn{nodename}, a legitimate but rare use case, leading to an incorrect inference by \syzdescribe.
Moreover, it fails to analyze the command value for \CodeIn{ioctl}, the interface to interact with the device.
This is because the command value undergoes a modification in the code, \CodeIn{cmd = \_IOC\_NR(command)}, where \CodeIn{command} is from users. Such scenarios are not accounted for by \syzdescribe, which erroneously uses \CodeIn{cmd} as the command value in its generated descriptions, as shown in Figure~\ref{fig:dm-ctl-example-c}.

\parabf{Key insight.}
Can we automate and improve the learning of various rules for generating high-quality specifications from the codebase with minimal effort?
We address this question based on the insight that modern Large Language Models (\llm{s})~\cite{gpt4,chatgpt,starcoder,codellama,codex,feng2020codebert,wei2023magicoder} are pre-trained on vast datasets, including kernel codebases, documentation, {and real-world syscall use cases.}{
The Linux kernel's extensive history and associated wealth of discussions, tutorials, and documentation further enrich this training data.}
Consequently, LLMs have likely been exposed to and potentially learned a wide range of information about syscall specifications.
This superior knowledge base makes them adept at analyzing kernel source code, even for atypically implemented syscalls, and generating \emph{high-quality, readable} specifications.

Utilizing \llm{s} as shown in Figure~\ref{fig:intro}, we can automate the process of inferring rules for mapping codebase content to syscall specifications and tailor these rules to be more general and adaptable to diverse cases.
Additionally, this approach eliminates the need for hard-coding rules within complex static analysis tools since \llm{s} inherently can analyze code, which significantly eases the process of adapting to evolving changes within the kernel codebase.
Returning to the case of the device mapper driver, \llm{s} demonstrate their capability to accurately infer broader rules.
They recognize that \CodeIn{.nodename} should be used as the device name when it is set and can identify modifications made to the command value.
Consequently, in our experiments, the specification generated by \llm{s} for the device mapper (Figure~\ref{fig:dm-ctl-example-d}) is not only correct but also more complete compared to those produced by \syzdescribe.
Impressively, this specification inferred by \llm{s} contributes to the discovery of 3 new bugs for this driver, 2 assigned with CVEs.

Building on the insight discussed above, we introduce \tech, the first approach to fully automate syscall specification generation by using Large Language Models (\llm{s}), focusing on kernel drivers and sockets.
The key idea of \tech is to employ \llm{s} for automating and enhancing the rule inference process, aimed at synthesizing high-quality syscall descriptions from their source code.
Taking the located operation handler as input, \tech recovers the identifier value (\eg device name or command value), type structure, and dependency for the syscalls related to the handler.
To this end, \tech iteratively applies \llm{s} to analyze the relevant source code and indicate the missing but essential information for inference, which will be analyzed in the next iteration.
Afterward, \tech validates and repairs the generated specifications by consulting \llm{s} with the error messages encountered.

Our contributions are summarized below:
\begin{itemize}[noitemsep, leftmargin=15pt, topsep=2pt]
    \item %
    We propose the first \emph{automated} approach to leveraging the potential of \llm{s} for kernel fuzzing.
    Moreover, different from existing LLM-based fuzzing work~\cite{titanfuzz,fuzz4all,yang2023white}, our approach goes beyond merely generating test inputs; we synthesize components of the fuzzing framework to integrate \llm{s} with matured frameworks developed for years, opening a new dimension for LLM-based fuzzing.
    \item We implement \tech to infer syscall specifications with a novel iterative strategy and further repair the descriptions with the validation feedback.
    Our artifact is available at \href{https://github.com/ise-uiuc/KernelGPT}{https://github.com/ise-uiuc/KernelGPT}.
    \item We evaluate \tech in generating new specifications to detect bugs and for the existing drivers and sockets to compare against state-of-the-art baselines, \syzdescribe and \syzkaller.
    Our experimental results show that \tech can generate more new and valid syscall descriptions and achieve higher coverage than baselines.
    \item \tech has already detected \numNewBug previously unknown bugs, with \numFix fixed and \numCVE CVE assignments, in the upstream Linux kernel.
    Notably, a number of specifications generated by \tech are merged into \syzkaller, following a request from its development team.
\end{itemize}

\section{Background}

\subsection{Kernel Fuzzing}

\parabf{OS Kernel Bugs.}
An OS kernel provides userspace applications with key functionalities, such as virtual memory, file system, networking, and access to devices.
To protect the safety of all applications and users, interactions between userspace and kernel are confined to a well-defined system call interfaces (\textbf{syscall}), \eg the POSIX standard. 
Kernel bugs that can be triggered through the syscall interface pose a significant risk since the interface is easily accessible to attackers.
Therefore, detecting bugs through the syscall interface has been an important direction of kernel security.
In this work, we focus on detecting kernel bugs through the Linux kernel system call interface.

\parabf{Device driver and socket.} Device drivers and sockets are the most complex and important components in the Linux kernel, comprising 41.6\% and 27\% of the LoC, respectively~\cite{bursey2024syzretrospector}. Due to the diversity of devices and network protocols, the syscall for interacting with drivers and sockets is complex.

Drivers and sockets register syscall handlers that are invoked when corresponding syscalls are used. Device drivers communicate with hardware upon receiving syscalls. Figure~\ref{fig:dm-ctl-example-a} and Figure~\ref{fig:dm-ctl-example-b} show data structures used for registering drivers. The \CodeIn{nodename} field represents the device file name, and the \CodeIn{fops} field stores function pointers for custom handler functions. When a user calls \CodeIn{open} with the \CodeIn{nodename}, kernel invokes the \CodeIn{dm\_open} handler and associates the file descriptor with the driver. Subsequent syscalls with this file descriptor invoke corresponding handlers. Sockets register syscalls like \CodeIn{socket}, \CodeIn{recvfrom}, and \CodeIn{setsockopt}. Each driver and socket registers different handlers, requiring unique specifications for effective fuzzing.

\parabf{Generic syscalls.}
While drivers and sockets can register syscall handlers, only a limited number of syscalls can be registered, and they may not cover all necessary operations. Generic syscalls like \textbf{\CodeIn{ioctl}} and \textbf{\CodeIn{setsockopt}} are heavily used. They have numeric parameters (identifier values) and an untyped pointer parameter. The numeric parameter identifies the operation, and the untyped pointer is cast to the required data structure. For example, to get the list of \CodeIn{dm} device names, an application calls \CodeIn{ioctl} with the \CodeIn{DM\_LIST\_DEVICES} macro and a pointer to \CodeIn{struct dm\_ioctl}. Sockets follow a similar pattern for programming \CodeIn{setsockopt} handlers.

The extensive usage of \emph{syscall handlers} and \emph{generic syscalls} turns a handful of syscalls into thousands of different syscalls, exposing a large attack surface.
Even worse, the implementations are scattered across a wide array of drivers and sockets.
This makes reasoning, analyzing, and testing the device drivers and sockets particularly challenging.

\begin{figure}[t]
    \includegraphics[width=0.9\columnwidth]{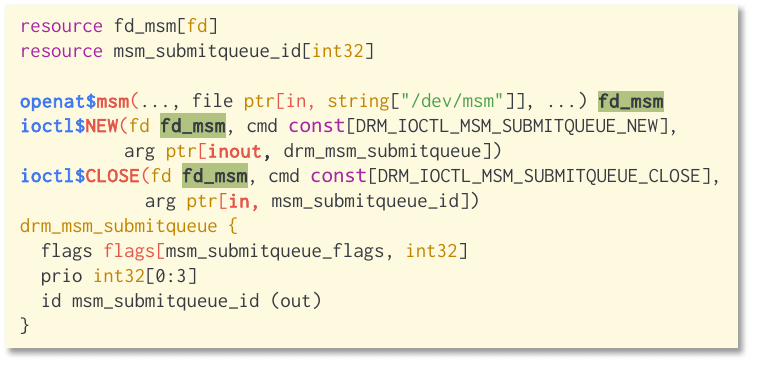}
    \caption{Specification for the MSM driver in \syzlang}
    \label{fig:syzlang}
\end{figure}

\parabf{Syscall Fuzzer.}
Among various methods for kernel bug detection~\cite{deviant, syzdescribe, syzdirect, syzvegas, syzgen, imf-paper, difuze, healer, moonshine, trinity, Triforce},
\syzkaller~\cite{syzkaller}, the state-of-the-art kernel fuzzer, has identified
thousands of kernel bugs. \syzkaller uses the syntax and semantics of
syscalls to generate diverse syscall sequences that can cover deep and diverse
code paths. To define the syntax and semantics of syscalls, \syzkaller provides
a domain-specific language, \syzlang\cite{syzlang}, to define \textbf{syscall
specifications} (or descriptions). Figure~\ref{fig:syzlang} are example
specifications of three syscalls for the MSM driver, showcasing \syzlang's
expressive power:

\begin{itemize}[noitemsep, leftmargin=15pt, topsep=2pt]
    \item \emph{Syntax}: The syntax of a system call is expressed by the definition of parameter types. The types \CodeIn{int32}, \CodeIn{string}, \CodeIn{ptr}, and the struct \CodeIn{rm\_msm\_submitqueue} allow \syzkaller to know how to structure the bytes for all the parameters.
    \item \emph{Semantically Valid Values}: Some parameters have specific value requirements. For example, the filename \CodeIn{"/dev/msm"} is the only valid file for the MSM driver.
    \item \emph{Inter-Syscall Dependency}: One syscall may depend on the output of another. The return value of \CodeIn{openat\$msm}, \CodeIn{fd\_msm}, is the same variable as the first inputs of the two \CodeIn{ioctl} syscalls, indicating their sequential execution.
    \item \emph{Intra-Syscall Dependency}: For generic syscalls like \CodeIn{ioctl}, the semantics of one parameter may depend on the value of another parameter.
    \syzkaller allows defining multiple specifications for the same syscall to express fine-grained semantics. 
    For example, for \CodeIn{ioctl\$NEW} and \CodeIn{ioctl\$CLOSE}, the types of the third parameter depend on the macro value of the second.
    \item \emph{Type Constraints}: Type definitions can incorporate semantic constraints. For example, in \CodeIn{rm\_msm\_submitqueue}, \CodeIn{[0:3]} represents the valid range of \CodeIn{prio}, and \CodeIn{(out)} denotes that \CodeIn{msm\_submitqueue\_id} is used as output.
\end{itemize}

While \syzlang supports more advanced features, their core functionalities are similar to those discussed. Effective specifications enable \syzkaller to reduce the search space by filtering out invalid syscall sequences. However, creating these specifications requires a deep understanding of syscall semantics, challenging both humans and automated tools.

\subsection{Specification Generation}
Specifications are typically manually written by \syzkaller and kernel developers, requiring deep expertise in kernel and the specific kernel module.
Thus, existing \syzkaller specifications only cover a subset of syscalls, especially for device drivers~\cite{bursey2024syzretrospector}.
As the kernel evolves, specifications can become out-of-date~\cite{bursey2024syzretrospector, syzdescribe}.
Automated specification generation has been desired for years~\cite{syzkaller-spec-gen-issue}, but faces several challenges.
First, discovering the interface of the vast number of operations implemented behind generic syscalls is challenging.
This involves inferring the correct operation identifier value (\eg device file name, socket domain, or \CodeIn{ioctl} command value), and then the corresponding data type for each unique operation.
Moreover, finding the dependencies among syscalls and data types is also key to reaching deep paths.
Failing to address these challenges would lead to inaccurate specifications, diminishing the effectiveness of a fuzzing campaign.
Aside from fuzzing performance, readability of the machine-generated specification is also crucial for human experts to validate and maintain~\cite{syzkaller-spec-gen-issue}.
Unreadable specifications could hide flaws that in turn hurt the effectiveness of fuzzing.

Several techniques for specification generation have been proposed, attempting to address some of the above challenges.
KSG~\cite{ksg} generates specifications by dynamically probing the kernel.
It first opens the devices (or sockets) existing in a booted environment and probes the kernel to detect their syscall handlers.
Then it infers the handler's parameter type through symbolic execution.
Relying on existing device files, KSG is unable to generate specifications for drivers that are not loaded in the kernel or require more setup steps.
In contrast, DIFUZE~\cite{difuze} and the state-of-the-art specification generation approach, \syzdescribe~\cite{syzdescribe}, both employ static analysis.
DIFUZE finds syscall handlers from a list of data structures used by common device registration functions.
\syzdescribe discovers syscall handlers by finding the kernel module initialization functions and tracing down to find the handler function pointers.
Both DIFUZE and \syzdescribe then conduct static analysis to identify the device file name, command value, and required parameter type.
Their static analysis models common implementation patterns \eg a switch case in a handler is likely invoking the corresponding sub-handlers based on the command value.

\subsection{Challenges and Opportunities}
\label{sec:motivation}

Existing static analysis approaches to specification generation face several limitations.

\parabf{\Challenge{1}: Incomplete modeling.}
Rule-based approaches struggle to capture the diversity of kernel code patterns, leading to limited coverage. Maintaining these rules is challenging and impractical.

\parabf{\Challenge{2}: Readability.}
Static analysis often generates specifications that are difficult for humans to understand, hindering validation and maintenance~\cite{syzkaller-spec-gen-issue}.

\parabf{\Challenge{3}: Textual comprehension.}
These tools struggle to infer specifications from textual information, such as comments, limiting their ability to capture the underlying meaning and intent of syscall behavior.

\parabf{\Solution{1}: Leveraging \llm{s}.}
To address the limitations, we propose a novel approach leveraging the strengths of \llm{s}:
\begin{itemize}[noitemsep, leftmargin=15pt, topsep=2pt]
    \item 
\emph{Mitigating \Challenge{1}}:
\llm{s} are pre-trained on extensive codebases, enabling them to handle a broader range of cases more effectively than static analysis rules.
\item
\emph{Mitigating \Challenge{2}}:
\llm{s} can generate descriptive and human-readable names within specifications based on code, enhancing readability and maintenance.
\item
\emph{Mitigating \Challenge{3}}:
\llm{s} excel in interpreting textual information, producing specifications that capture the underlying meaning and intent of syscall behaviors.

\end{itemize}

While harnessing the potential of \llm{s}, we must design strategies to mitigate their inherent limitations, such as context size restrictions and hallucinations~\cite{ji2023survey}. To achieve this, we 1) incorporate syz-lang knowledge through few-shot prompting, 2) develop a novel iterative multi-stage prompting approach, and 3) leverage off-the-shelf validation tools for debugging.
More details will be discussed in \S~\ref{sec:design}.

\begin{figure*}[t]
    \includegraphics[width=0.9\linewidth]{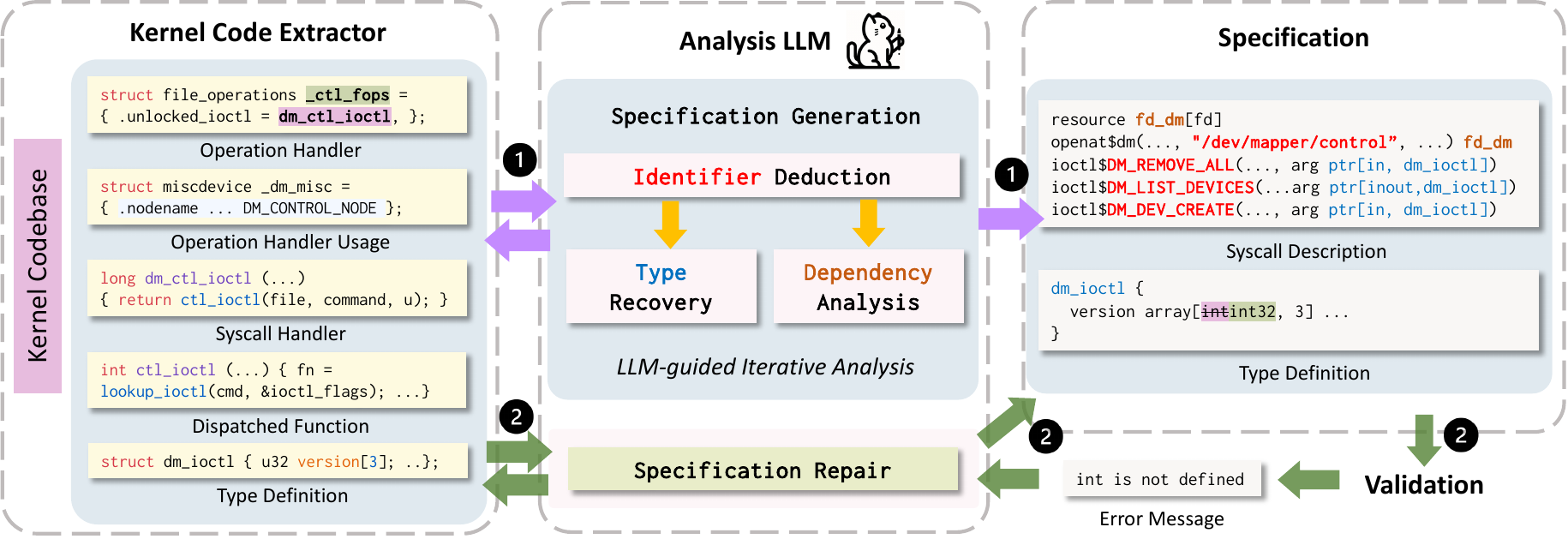}
    \caption{Overview of \tech}
    \label{fig:overview}
\end{figure*}

\section{Design}

\label{sec:design}

Figure~\ref{fig:overview} presents the overview workflow of \tech, which utilizes a code extractor and an analysis \llm{} (\S~\ref{sec:impl}) to fully automatically generate specifications for kernel fuzzing.

\tech takes the kernel codebase and located operation handlers as input and operates through two automated phases: Specification Generation \circled{1}, and Specification Validation and Repair \circled{2}.
Initially, \tech determines the identifier values (\S~\ref{subsec:id-deduction}), argument types (\S~\ref{subsec:type}), and dependencies (\S~\ref{subsec:dependency}) for describing the syscalls associated with the given operation handler.
In doing so, \tech utilizes the relevant source code from the kernel codebase to guide the \llm{s} in their analysis in a novel iterative way.
If essential information for inference is missing, the analysis \llm{} is instructed to indicate what additional information is required, which is then gathered and presented for analysis in the following step (\S~\ref{subsec:spec-generation} \circled{1}).
Subsequently, \tech validates the generated specifications. If errors are found, it attempts to repair the descriptions by consulting the \llm{s} with the error messages (\S~\ref{subsec:spec-repair} \circled{2}).

\subsection{Specification Generation}
\label{subsec:spec-generation}

\tech generates the specifications for syscalls by leveraging \llm{s} to analyze
the implementation source code of the syscalls. Initially, we identify the
syscall handler functions (e.g. \CodeIn{ioctl} and \CodeIn{setsockopt}) from
their respective operation handlers.
We segment the specification generation process into three stages: identifier deduction, type recovery, and dependency analysis.
This pipeline enables \llm{s} to focus on one specific aspect at each stage and
avoid misleading information from irrelevant code snippets. In each stage,
we utilize in-context few-shot prompting~\cite{fewshot} to enhance \llm{s}'
comprehension of the task and formalize output.

\begin{algorithm}[t]
\small
\caption{\llm-Guided Iterative Analysis}
\label{algo:analyze}
\DontPrintSemicolon
\SetKwProg{Fn}{Function}{:}{}

\SetKwData{relatedsrc}{relatedCode}
\SetKwData{usage}{usageInfo}
\SetKwData{step}{step}
\SetKwData{result}{result}
\SetKwData{success}{result}
\SetKwData{unknown}{unknown}
\SetKwData{id}{id}
\SetKwData{relatedfunc}{funcs}
\SetKwData{relatedtype}{types}
\SetKwData{res}{res}
\SetKwData{prompt}{prompt}

\SetKwData{maxiter}{MAX\_ITER}
\SetKwFunction{QueryLLM}{\textbf{QueryLLM}}
\SetKwFunction{FindCode}{\textbf{ExtractCode}}
\SetKwFunction{update}{\textbf{Update}}
\SetKwFunction{Analyze}{\textbf{Analyze}}
\SetKwFunction{GenPrompt}{\textbf{GenPrompt}}

\Fn{\Analyze{\relatedsrc, \usage, \step}}{
\If{\step > \maxiter}{
    \Return{$\emptyset$}\; \label{line:max-iter}
}
\textit{\scriptsize\color{applegreen}\# Prepare the prompt with few-shot examples}\;
\prompt $\leftarrow$ \GenPrompt(\relatedsrc, \usage) \; \label{line:prompt-gen}
\textit{\scriptsize\color{applegreen}\# Query \llm{} to analyze the source code and identify the unknown}\;
\success, \unknown $\leftarrow$ \QueryLLM(\prompt)\; \label{line:query}
\For{$(\id, \usage) \in \unknown$}{\label{algo:sample-start}
    \textit{\scriptsize\color{applegreen}\# Extract the code based on the unknown identifier}\;
    \relatedsrc $\leftarrow$ \FindCode(\id)\label{line:extract}\;
    \textit{\scriptsize\color{applegreen}\# Recursively analyze the missing source code}\;
    \res $\leftarrow$ \Analyze(\relatedsrc, \usage, $\step+1$)\; \label{line:next-step}
    \textit{\scriptsize\color{applegreen}\# Update with the result analyzed for the unknown}\;
    \update(\success, \res)\;
}\label{algo:sample-end}
\Return{$\success$}\;
}

\end{algorithm}

\parabf{Iterative analysis.}
All three stages follow an iterative analysis paradigm. The motivation for
this iterative design is two-fold. First, even though state-of-the-art \llm{s}
like \gpt{4} can support long context size up to 128K~\cite{gpt4-turbo}, this
size is still not enough to provide the entire source code related to a syscall. Second, the goal of specification generation is to deduce the
identifier value, type structure, and dependency for the syscall.
However, not all code or helper functions within the syscall handler are directly relevant to this goal. Consequently, we allow \llm{s} to identify the pertinent source code for the current goal.

The pipeline of the iterative analysis is shown in Algorithm~\ref{algo:analyze}.
First, we generate a few-shot example prompt with the source code related to the
target syscall and its usage information (Line~\ref{line:prompt-gen}). Then, we
query \llm{s} to infer the descriptions and pinpoint the unknown targets
(functions or types; Line~\ref{line:query}). Such unknown functions/types are
the ones that are missing in the provided prompt yet are essential for the
inference. 
\CodeIn{unknown} set. Then, for each unknown target, we extract its source code by using its identifier.
This code is then fed to
\llm{s} along with its usage information for further analysis
(Line~\ref{line:next-step}). This iterative algorithm continues until there is
no unknown target or the iteration reaches a predefined threshold
\CodeIn{MAX\_ITER} (Line~\ref{line:max-iter}). Note that the entire analysis
process is \emph{fully automated} and requires no human intervention. The unknown
information provided by \llm{s} is used to guide the analysis process.
Next, we demonstrate in detail how each stage employs this iterative algorithm.

\subsubsection{Identifier Deduction}\label{subsec:id-deduction}
The first step
of specification generation is to deduce the identifier value of the syscall. To
achieve this goal, we utilize the iterative strategy described in
Algorithm~\ref{algo:analyze} to analyze the syscall-related source code. The
expected output from \llm{s} (the output of \CodeIn{QueryLLM}) is the set of
successfully inferred identifier values (\CodeIn{result}). If the logic for checking identifier values is delegated to another function not presented to \llm{s}, we instruct \llm{s} to list the name and invocation details of this ``missing'' dispatched function (the variable \CodeIn{unknown}). Besides, we also include code snippets that reference the command variables. If \llm{s} identifies any unknown identifier values, \tech proceeds to analyze the newly identified dispatched function, incorporating their usage information from the previous step. In essence, the output from the \CodeIn{unknown} of the previous step serves as a reference for guiding subsequent steps.

Compared to traditional static analysis, \llm{s} displays great potential to handle a wider range of scenarios in identifier value inference.
To help with the \llm{s} understanding, we provide a few-shot example within the prompt (\CodeIn{GenPrompt}). These examples serve as guides for \llm{s} to improve their reasoning and deduce the identifier values more effectively.

\subsubsection{Type Recovery}\label{subsec:type}
Following identifier value inference, the next stage is to analyze the argument type structure for each identifier value. Leveraging information from the identifier deduction stage, we extract related functions and present them to the LLM to identify argument types. If type determination logic is delegated to other functions, we continue the analysis in subsequent steps, using the new information to guide the process.
After determining the argument types, \tech generates descriptions for these types.
By retrieving type definition source code from the \kernel codebase and feeding it to the LLM, we obtain \syzkaller descriptions. If nested types are encountered, they are marked as \CodeIn{unknown} for further analysis in following steps.

\begin{figure}[t]
    \centering
    \captionsetup{belowskip=-6pt}
    \includegraphics[width=0.9\columnwidth]{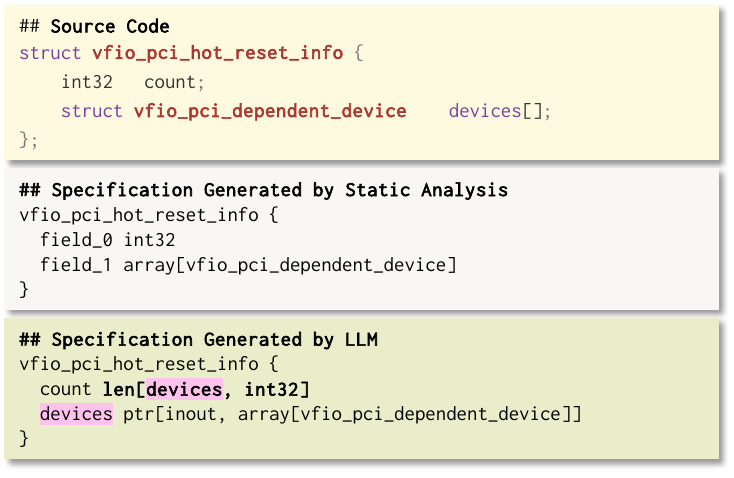}
    \caption{Type definitions from static analysis and \llm}
    \label{fig:type-example}
\end{figure}

While static analysis can recover type definitions from source code, it is difficult to infer the semantic relationships between nested types, particularly within \CodeIn{struct} and \CodeIn{union} definitions.
For instance, consider the field \CodeIn{count} within the structure in Figure~\ref{fig:type-example}. \CodeIn{count} represents the number of elements within another field, \CodeIn{devices}.
Traditional static analysis tools struggle to capture this inherent relationship and treat these fields independently.
In contrast, \llm{s} can understand the semantic connections between different fields or types within nested structures. \tech leverages this capability to generate descriptions that capture these relationships. As depicted in Figure~\ref{fig:type-example}, the description produced by \tech is \CodeIn{count len[devices]}. The generated description can effectively capture the semantic connection between \CodeIn{count} and the number of elements in \CodeIn{devices}.

\subsubsection{Dependency Analysis}\label{subsec:dependency}
Finally, we need to analyze the dependencies between syscalls. Specifically, the dependency means whether another syscall (or operation handler) relies on the return value of the current syscall. To achieve this goal, we leverage \llm{s} to identify if the return value could be a resource (\eg file descriptor) of another operation handler.
\tech extracts the source code of the relevant functions and presents the code to \llm{s}.
Notably, the return value relevant functions are marked by \llm{s} themselves in the first stage (\S~\ref{subsec:id-deduction}).
Suppose the return value can be used by other syscalls, \llm{s} identifies them. If the logic for dependency analysis resides in other functions, \llm{s} marks them as \CodeIn{unknown}, and \tech utilizes the new information to continue the analysis in subsequent steps.

\subsection{Specification Validation and Repair}
\label{subsec:spec-repair}

In this phase, inspired by recent work on \llm-based program repair~\cite{chatrepair}, our goal is to validate the specifications generated by \tech and automatically repair the invalid ones.
This is because \llm{s} may occasionally make mistakes during the description generation process.
To address potential inaccuracies, we employ off-the-shelf validation tools. These tools analyze the specifications and provide error messages if discrepancies are found.
Initially, \tech uses the error messages from them to pinpoint inaccuracies in specific descriptions, effectively matching each error message to its corresponding description.
Then, for those descriptions identified with errors, \tech queries \llm{s} for correction, guided by few-shot examples.
This process involves supplying \llm{s} with the incorrect description, the associated error messages, and relevant source code from the kernel codebase to repair.
\llm{s} are then expected to output the correct descriptions.

\revision{Existing validation tools are limited in their ability to detect semantic errors~\cite{syzlang}, primarily focusing on syntax validation and simple semantic checks. For example, runtime validation of semantic correctness remains a significant challenge, which is why current syscall specification generation approaches do not incorporate it~\cite{syzdescribe,difuze}. To ensure a more thorough evaluation, we manually examined the generated specifications (\S~\ref{subsec:spec-manual-eval}), demonstrating that \tech successfully synthesizes semantically correct specifications.}

\section{Implementation}
\label{sec:impl}
\parabf{Target.}
While our approach is general to various syscalls, \tech targets those for kernel drivers and sockets, given the fact that they constitute about 70\% LoC in the kernel~\cite{bursey2024syzretrospector}.
For drivers, we focus on the critical \CodeIn{ioctl} syscall, in addition to initialization syscalls such as \CodeIn{openat} and \CodeIn{sys\_open\_dev}.
For sockets, we extend our support beyond \CodeIn{socket} and \CodeIn{ioctl} to include syscalls like \CodeIn{bind}, \CodeIn{connect}, \CodeIn{accept}, \CodeIn{poll}, \CodeIn{sendto}, \CodeIn{recvfrom}, \CodeIn{setsockopt}, and \CodeIn{getsockopt}.

\parabf{Source code extractor.}
It is implemented using the \llvm toolchain~\cite{llvm} and parses the kernel codebase to: 
\begin{itemize}[noitemsep, leftmargin=15pt, topsep=2pt]
    \item \emph{Driver and Socket Operation Handler Extraction.} The extractor employs \emph{simple yet general} pattern matching to pinpoint driver and socket operation handlers. These are then prepared as inputs for \tech, extracted with their corresponding usage locations.
    More specifically, we search for initialization instances of the \CodeIn{ioctl} or \CodeIn{unlocked\_ioctl} fields within the operation handlers.
    For instance, the device mapper driver shown in Figure~\ref{fig:dm-ctl-example-a} initiates the \CodeIn{unlocked\_ioctl} field in the structure \CodeIn{\_ctl\_fops} by using \CodeIn{dm\_ctl\_ioctl} function.
    We label \CodeIn{\_ctl\_fops} as the device operation handler and extract \CodeIn{dm\_ctl\_ioctl} to generate specifications for \CodeIn{ioctl} syscalls.
    \tech focuses on inference from source code to descriptions, so we use a straightforward pattern-searching method to find device and socket operations.
    \item \emph{Kernel Definition Extraction.} The extractor compiles all definitions of \CodeIn{function}, \CodeIn{struct}, \CodeIn{union}, and \CodeIn{enum} found within the kernel. These definitions are used as guidance for \llm{s} in the specification generation and repair processes, provided when \llm{s} indicate their necessity (\FindCode function in Algorithm~\ref{algo:analyze}).
\end{itemize}

\parabf{Analysis \llm.}
While our approach is general and independent of the specific \llm{s} used, our tool, \tech, is constructed atop \gpt{-4}~\cite{gpt4}. At each step, we utilize the OpenAI APIs to query \gpt{-4}, with a low-temperature of \CodeIn{0.1}.
We set the stopping criteria for analysis as 5 by default.

\begin{figure}[t]
    \centering
    \includegraphics[width=0.9\columnwidth]{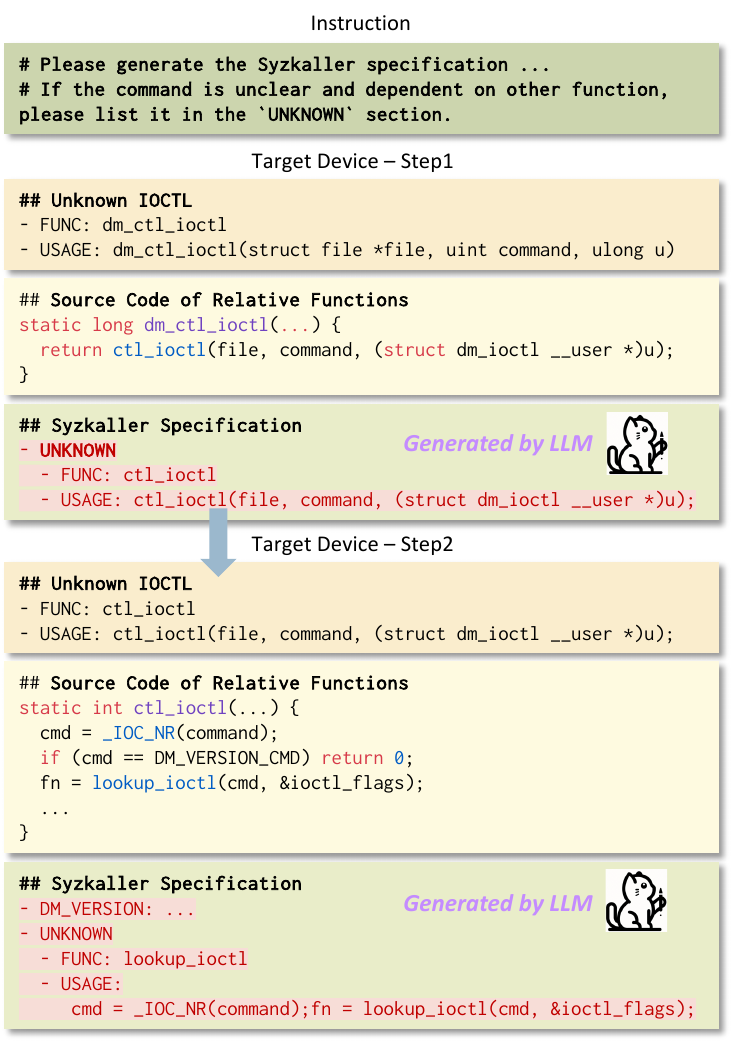}
    \caption{Iterative prompt for identifier deduction}
    \label{fig:step1}
\end{figure}

\parabf{Iterative analysis.}
We design a structured prompt template to facilitate interaction with \llm{s} for the kernel code analysis.
For instance, Figure~\ref{fig:step1} presents the first two steps of identifier deduction (\S~\ref{subsec:id-deduction}) for the device mapper driver.
The \CodeIn{dm\_ctl\_ioctl} handler, registered as the \CodeIn{ioctl} handler, offloads its entire functionality to another function, \CodeIn{ctl\_ioctl}.
As a result, after examining \CodeIn{dm\_ctl\_ioctl}, \llm{s} are unable to deduce any identifier values and designate \CodeIn{ctl\_ioctl} as the absent function.
Then \tech extracts the source code for \CodeIn{ctl\_ioctl} and, together with the unknown information returned by the first step, re-queries \llm{s}.
In the second round, \llm{s} successfully identifies one identifier value, \CodeIn{DM\_VERSION}, while other values related to the function \CodeIn{lookup\_ioctl} remain undetermined.
Thus, \llm{s} report \CodeIn{DM\_VERSION} as one identifier value and \CodeIn{lookup\_ioctl} as the missing function, which will be analyzed in the next step to infer more identifier values.

\revision{We set the default value of \CodeIn{MAX\_ITER} to 5 (Line~\ref{line:max-iter}, Algorithm~\ref{algo:analyze}). For efficiency, our implementation caches and reuses results from previously explored paths. With these configurations, we observed no termination issues throughout our experiments.}

\parabf{Specification generation.}
Rather than generating descriptions for all possible drivers and sockets, we focus on those not covered by existing specifications, which are often less thoroughly tested.
We select drivers and sockets activated in our configuration, excluding ones used for debugging (e.g., \CodeIn{/dev/gup\_test}) or requiring specific hardware/architecture, as testing them would be meaningless or impractical.
\revision{This filtering process is largely automated: debug drivers are easily identified by their \CodeIn{\_test} suffix, and hardware-specific drivers can be filtered by focusing only on bootable modules.}

\parabf{Validation.}
We leverage two tools in \syzkaller, \CodeIn{syz-extract} and \CodeIn{syz-generate} to validate the generated specifications, which can detect many types of errors, including issues such as undefined types, wrong macro names, unmatched dependencies, and more.

\section{Evaluation}

We conduct an extensive evaluation on a workstation with 96 cores and 512 GB RAM, running Ubuntu 20.04.5 LTS.
We selected the Linux kernel version 6.7 (\CodeIn{d2f51b}) as our target.
Following prior work~\cite{syzdescribe}, we use the \CodeIn{allyesconfig} kernel configuration for specification generation, but use the \CodeIn{syzbot}~\cite{syzbot} configuration from Google to build a bootable kernel for evaluation.
We use the \syzkaller setting for fuzzing, with 4 QEMU instances, each utilizing 2 CPU cores.
For baselines,
we choose \syzdescribe~\cite{syzdescribe}, the state-of-the-art syscall specification generation approach, and existing \syzkaller~\cite{syzkaller} specifications, crafted by human experts.

\subsection{Specification Generation for Missing}
\label{subsec:sepc-gen-eval}

\begin{table}[t]\centering
\caption{Specifications for driver/socket handlers}
\label{tab:overall}
\footnotesize
\begin{tabular}{lrrrrr}\toprule
& & &\syzdescribe &\tech \\\cmidrule{4-5}
&\# Total &\# Incomplete &\# Valid &\# Valid (Fixed)\\\midrule
Driver &278 &75 &20 &70 (30) \\
Socket &81 &66 &N/A &57 (12) \\ \midrule
Total &359 &141 &20 &127 (42) \\
\bottomrule
\end{tabular}
\end{table}

\begin{figure}[!t]
    \centering
    \begin{subfigure}[b]{0.48\linewidth} %
        \includegraphics[width=\linewidth]{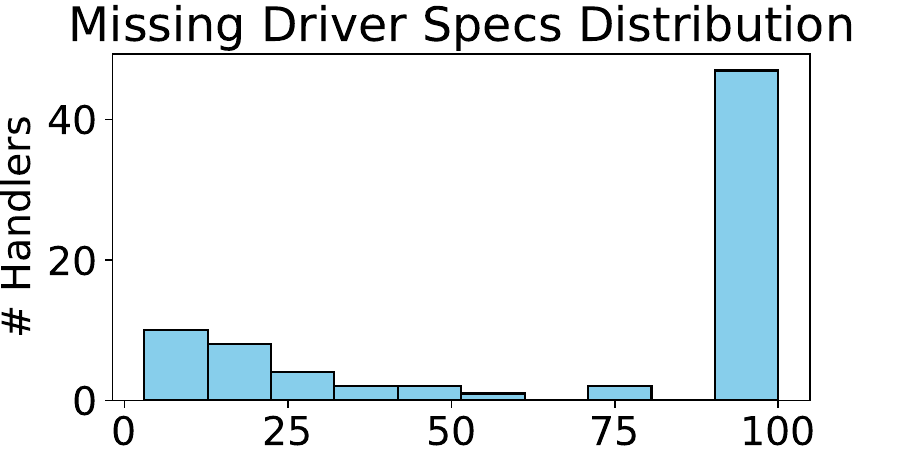}
        \label{fig:driver-new}
    \end{subfigure}
    \hfill %
    \begin{subfigure}[b]{0.48\linewidth} %
        \includegraphics[width=\linewidth]{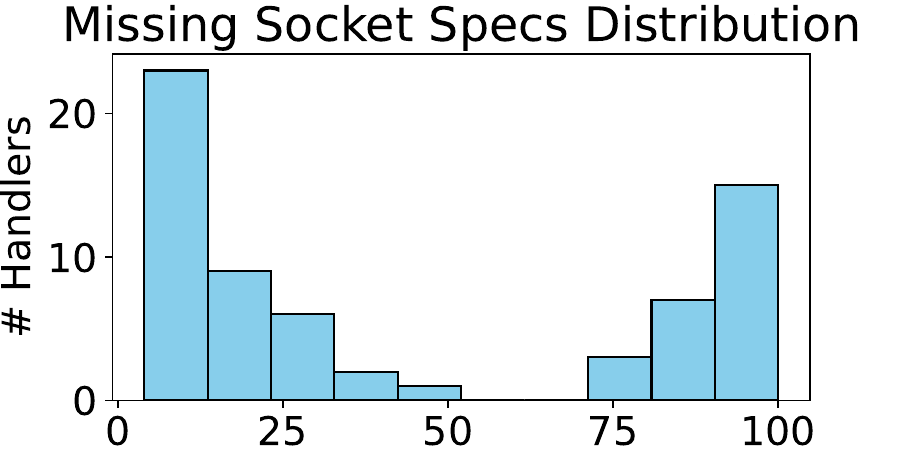}
        \label{fig:socket-new}
    \end{subfigure}
    \caption{Missing specification distribution}
    \label{fig:newspec}
\end{figure}

Under the \textbf{\CodeIn{allyesconfig}} option,
\tech scans \emph{\totalDriver driver} operation and \emph{\totalSocket socket} operation handlers.
Out of the them, \indevDriver and \indevSocket are respectively \textbf{loaded} under the \textbf{\CodeIn{syzbot}} option.
For these {loaded} driver and socket operation handlers, we generated
\emph{missing} descriptions, as presented in Table~\ref{tab:overall}.
After analysis, \missingDriver driver and \missingSocket socket operation handlers are missing one or more syscall descriptions (Column ``\# Incomplete'').
Figure~\ref{fig:newspec} presents a histogram where the x-axis represents the percentage of missing syscall specifications, and the y-axis shows the count of handlers at each percentage level. 
We note that \syzkaller does not have any description for many of these driver handlers (45 out of \missingDriver, or 60\%).
Plus, 22 socket handlers lack descriptions for over 80\% of their syscalls.
These findings underscore the insufficient specification of some drivers and sockets for effective testing,
highlighting the need to synthesize additional specifications.

\subsubsection{Statistics of New Specifications}
Among the \missingDriver driver and \missingSocket socket handlers with missing syscall descriptions, \validDriverNoRep and \validSocketNoRep are directly validated as correct (\S~\ref{subsec:spec-generation}), and additional \repairDriver and \repairSocket are successfully repaired (\S~\ref{subsec:spec-repair}).
Hence, \tech successfully generates specifications for \validDriver (93\%) and \validSocket (86\%) of the driver and socket handlers with missing syscalls, demonstrating the effectiveness of \tech for specification generation and repair.
By contrast, the state-of-the-art syscall specification generation approach, \syzdescribe~\cite{syzdescribe}, is limited to generating specifications for only 20 (27\%) inadequately described driver handlers and cannot analyze socket handlers at all.

\begin{table}[t]\centering
\caption{Newly generated syscall descriptions}
\label{tab:new-desc}
\footnotesize
\begin{tabular}{l|rr|rrr}\toprule
&\multicolumn{2}{c}{\syzdescribe} &\multicolumn{2}{c}{\tech} \\\cmidrule{2-5}
&\# Syscalls &\# Types &\# Syscalls &\# Types \\\midrule
Driver &146 &168 &288 &170 \\
Socket &N/A &N/A &244 &124 \\ \midrule
Total &146 &168 &532 &294 \\
\bottomrule
\end{tabular}
\end{table}

The descriptions generated by \tech for these handlers includes \numNewSpec (\newPercent) \emph{new} syscalls, in addition to the \numSyzSpec existing syscalls described by \syzkaller, as shown in Table~\ref{tab:new-desc}.
This also includes additional \numNewType descriptions for new types employed within these syscall specifications.
By contrast, \syzdescribe has only \numDescNewSpec (\numDescNewPercent) new syscall descriptions and \numDescNewType new type definitions for the drivers.
Again, \syzdescribe lacks support for analyzing sockets, resulting in ``N/A'' entries in the corresponding table sections.

\revision{\tech takes 4.7 hours to generate specifications for these 532 syscalls, more efficient than \syzdescribe, which requires 3.8 hours for just 146 syscalls. It processes approximately 5.56 million input tokens and generates 400,000 output tokens, with an average of 2,630 input and 189 output tokens per prompt. The total cost of ~\$34 is negligible considering these specifications guide extensive fuzzing campaigns that typically run for days or weeks.}

\llm{}-generated specifications offer unmatched readability compared to previous techniques.
While existing methods often use random numbers for syscall names, file descriptor names, and struct field names, \tech leverages \llm{}'s ability to generate meaningful names, closely resembling expert-written specifications.
For example, we have been requested by \syzkaller developers to upstream our generated specification for the CEC driver after we reported several bugs in the driver.
Our generated specification, covering 12 syscalls and 10 structs/unions with 47 fields in total, was merged into \syzkaller with only one word changed manually~\cite{kernelgpt-spec}.
In contrast, a previous attempt to merge a number of \syzdescribe generated specifications led to lengthy code review discussions and was not merged~\cite{syzdescribe-pr}.
A quote from \syzkaller developers also highlights the importance of specification readability:\textit{``Lots of automated descriptions that I saw are unreadable ...
When you start digging they turn out to be bad in some way, but discovering that is extremely hard, it should be easy (e.g. literal constant names)''}~\cite{syzkaller-spec-gen-issue}.

\begin{table}[t]\centering
\caption{Overall effectiveness of \tech (3 rep.)}
\label{tab:allcov}
\footnotesize
\begin{tabular}{l|rcr}\toprule
&Cov &Unique Cov &Crash \\\midrule
\syzkaller &204,923 &- &16.0 \\
\syzkaller+ \syzdescribe &201,634 &14,585 &13.7 \\
\syzkaller+ \tech &209,673 &20,472 &17.7 \\
\bottomrule
\end{tabular}
\end{table}

\subsubsection{Coverage Improvement by New Specifications}
To show the effectiveness of the new specifications synthesized by \tech for kernel fuzzing, we integrate them with the existing \syzkaller specifications, resulting in a combined suite (\syzkaller + \tech).
We conduct a 24-hour fuzzing session (192 CPU hours).
For comparison, we also run the original \syzkaller and a combined suite of \syzkaller with the descriptions generated by \syzdescribe (\syzkaller + \syzdescribe), each under identical conditions and with three repetitions.
The results are depicted in Table~\ref{tab:allcov}, where the \syzkaller + \tech suite covers 4,750 and 8,039 more \revision{basic blocks} than the original \syzkaller and the \syzdescribe integration, respectively.
Additionally, \tech adds 20,742 \emph{unique} \revision{basic blocks} to \syzkaller's coverage, in contrast to \syzdescribe's contribution of 14,585 additional unique \revision{blocks}.
These results underscore the contribution of \tech-generated missing descriptions to enhancing coverage in kernel fuzzing.
\begin{table*}[t]\centering
\caption{New bugs detected by \tech}
\label{tab:bug}
\footnotesize
\begin{tabular}{l|ccc|c|ccc}\toprule
\textbf{Crash with new specs} &\textbf{New} &\textbf{Confirmed} &\textbf{Fixed} &\textbf{CVE} &\textbf{\syzkaller} &\textbf{\syzdescribe} \\\midrule
kmalloc bug in ctl\_ioctl &\textcolor{darkgreen}{\textbf{\checkmark}} &\textcolor{darkgreen}{\textbf{\checkmark}} &\textcolor{darkgreen}{\textbf{\checkmark}} &CVE-2024-23851 &\textcolor{red}{\textbf{$\times$}} &\textcolor{red}{\textbf{$\times$}} \\
kmalloc bug in dm\_table\_create &\textcolor{darkgreen}{\textbf{\checkmark}} &\textcolor{darkgreen}{\textbf{\checkmark}} &\textcolor{darkgreen}{\textbf{\checkmark}} &CVE-2023-52429 &\textcolor{red}{\textbf{$\times$}} &\textcolor{red}{\textbf{$\times$}} \\
KASAN: slab-use-after-free Read in cec\_queue\_msg\_fh &\textcolor{darkgreen}{\textbf{\checkmark}} &\textcolor{darkgreen}{\textbf{\checkmark}} &\textcolor{darkgreen}{\textbf{\checkmark}} &CVE-2024-23848 &\textcolor{red}{\textbf{$\times$}} &\textcolor{red}{\textbf{$\times$}} \\
ODEBUG bug in cec\_transmit\_msg\_fh &\textcolor{darkgreen}{\textbf{\checkmark}} &\textcolor{darkgreen}{\textbf{\checkmark}} &\textcolor{darkgreen}{\textbf{\checkmark}} & &\textcolor{red}{\textbf{$\times$}} &\textcolor{red}{\textbf{$\times$}} \\
WARNING in cec\_data\_cancel &\textcolor{darkgreen}{\textbf{\checkmark}} &\textcolor{darkgreen}{\textbf{\checkmark}} &\textcolor{darkgreen}{\textbf{\checkmark}} & &\textcolor{red}{\textbf{$\times$}} &\textcolor{red}{\textbf{$\times$}} \\
INFO: task hung in cec\_claim\_log\_addrs &\textcolor{darkgreen}{\textbf{\checkmark}} &\textcolor{darkgreen}{\textbf{\checkmark}} & & &\textcolor{red}{\textbf{$\times$}} &\textcolor{red}{\textbf{$\times$}} \\
general protection fault in cec\_transmit\_done\_ts &\textcolor{darkgreen}{\textbf{\checkmark}} &\textcolor{darkgreen}{\textbf{\checkmark}} &\textcolor{darkgreen}{\textbf{\checkmark}} & &\textcolor{red}{\textbf{$\times$}} &\textcolor{red}{\textbf{$\times$}} \\
kernel BUG in btrfs\_get\_root\_ref &\textcolor{darkgreen}{\textbf{\checkmark}} &\textcolor{darkgreen}{\textbf{\checkmark}} &\textcolor{darkgreen}{\textbf{\checkmark}} &CVE-2024-23850 &\textcolor{red}{\textbf{$\times$}} &\textcolor{red}{\textbf{$\times$}} \\
general protection fault in btrfs\_update\_reloc\_root &\textcolor{darkgreen}{\textbf{\checkmark}} &\textcolor{darkgreen}{\textbf{\checkmark}} & & &\textcolor{red}{\textbf{$\times$}} &\textcolor{red}{\textbf{$\times$}} \\
zero-size vmalloc in ubi\_read\_volume\_table &\textcolor{darkgreen}{\textbf{\checkmark}} &\textcolor{darkgreen}{\textbf{\checkmark}} &\textcolor{darkgreen}{\textbf{\checkmark}} &CVE-2024-25739 &\textcolor{red}{\textbf{$\times$}} &\textcolor{red}{\textbf{$\times$}} \\
UBSAN: array-index-out-of-bounds in rds\_cmsg\_recv &\textcolor{darkgreen}{\textbf{\checkmark}} &\textcolor{darkgreen}{\textbf{\checkmark}} &\textcolor{darkgreen}{\textbf{\checkmark}} &CVE-2024-23849 &\textcolor{red}{\textbf{$\times$}} &\textcolor{red}{\textbf{$\times$}} \\
memory leak in ubi\_attach &\textcolor{darkgreen}{\textbf{\checkmark}} &\textcolor{darkgreen}{\textbf{\checkmark}} & &CVE-2024-25740 &\textcolor{red}{\textbf{$\times$}} &\textcolor{red}{\textbf{$\times$}} \\
memory leak in posix\_clock\_open &\textcolor{darkgreen}{\textbf{\checkmark}} &\textcolor{darkgreen}{\textbf{\checkmark}} &\textcolor{darkgreen}{\textbf{\checkmark}} &CVE-2024-26655 &\textcolor{red}{\textbf{$\times$}} &\textcolor{red}{\textbf{$\times$}} \\
memory leak in \_\_ip6\_append\_data &\textcolor{darkgreen}{\textbf{\checkmark}} &\textcolor{darkgreen}{\textbf{\checkmark}} & & &\textcolor{red}{\textbf{$\times$}} &\textcolor{red}{\textbf{$\times$}} \\
possible deadlock in dvb\_demux\_release &\textcolor{darkgreen}{\textbf{\checkmark}} & & & &\textcolor{red}{\textbf{$\times$}} &\textcolor{red}{\textbf{$\times$}} \\
INFO: task hung in \_\_rq\_qos\_throttle &\textcolor{darkgreen}{\textbf{\checkmark}} & & & &\textcolor{red}{\textbf{$\times$}} &\textcolor{red}{\textbf{$\times$}} \\
WARNING in usb\_ep\_queue &\textcolor{darkgreen}{\textbf{\checkmark}} &\textcolor{darkgreen}{\textbf{\checkmark}} & &CVE-2024-25741 &\textcolor{red}{\textbf{$\times$}} &\textcolor{red}{\textbf{$\times$}} \\
memory leak in dvb\_dmxdev\_add\_pid &\textcolor{darkgreen}{\textbf{\checkmark}} & \textcolor{darkgreen}{\textbf{\checkmark}}& & &\textcolor{red}{\textbf{$\times$}} &\textcolor{red}{\textbf{$\times$}} \\
memory leak in dvb\_dvr\_do\_ioctl &\textcolor{darkgreen}{\textbf{\checkmark}} & & & &\textcolor{red}{\textbf{$\times$}} &\textcolor{red}{\textbf{$\times$}} \\ 
general protection fault in dvb\_vb2\_expbuf &\textcolor{darkgreen}{\textbf{\checkmark}} & \textcolor{darkgreen}{\textbf{\checkmark}} & \textcolor{darkgreen}{\textbf{\checkmark}} & CVE-2024-50291 &\textcolor{red}{\textbf{$\times$}} &\textcolor{red}{\textbf{$\times$}} \\ 
general protection fault in cleanup\_mapped\_device &\textcolor{darkgreen}{\textbf{\checkmark}} & \textcolor{darkgreen}{\textbf{\checkmark}}&\textcolor{darkgreen}{\textbf{\checkmark}} & CVE-2024-50277 &\textcolor{red}{\textbf{$\times$}} &\textcolor{red}{\textbf{$\times$}} \\
WARNING in vb2\_core\_reqbufs &\textcolor{darkgreen}{\textbf{\checkmark}} & \textcolor{darkgreen}{\textbf{\checkmark}}& & &\textcolor{red}{\textbf{$\times$}} &\textcolor{red}{\textbf{$\times$}} \\
BUG: corrupted list in vep\_queue &\textcolor{darkgreen}{\textbf{\checkmark}} & \textcolor{darkgreen}{\textbf{\checkmark}}& & &\textcolor{red}{\textbf{$\times$}} &\textcolor{red}{\textbf{$\times$}} \\
divide error in uvc\_queue\_setup &\textcolor{darkgreen}{\textbf{\checkmark}} & \textcolor{darkgreen}{\textbf{\checkmark}}& & &\textcolor{red}{\textbf{$\times$}} &\textcolor{red}{\textbf{$\times$}} \\
\midrule
\textbf{Total} &\textbf{24} &\textbf{21} &\textbf{12} &\textbf{\numCVE} &\textbf{0} &\textbf{0} \\
\bottomrule
\end{tabular}
\end{table*}
\subsubsection{Correctness of New Specifications}
\label{subsec:spec-manual-eval}

To assess the semantic correctness of \tech-generated specifications, we manually examined the specifications for 45 drivers devoid of descriptions in \syzkaller (detailed in \S~\ref{subsec:sepc-gen-eval}), encompassing a total of 313 \CodeIn{IOCTL} syscall descriptions.
Primarily, we focused on syscalls that \tech overlooked, those with incorrect identifiers, and syscalls featuring erroneous types.

Regarding missing syscalls, we discovered that the majority of drivers (42/45, 93.3\%) did not omit any syscall. For the three drivers with missing syscall descriptions, their actual syscall handling was delegated to other functions, sometimes even multiple times. This underscores the challenges LLMs face in analyzing indirect function calls.
Moreover, we identified only 3 (0.9\%) syscalls out of 2 (4.4\%) drivers with incorrect identifier values. Upon closer inspection, we determined that modifications to identifier values, such as \CodeIn{if (DRM\_IOCTL\_NR(cmd) == DRM\_COMMAND)}, which checks the modified identifier value, and \CodeIn{DRM\_COMMAND} not being the true identifier value in this context, were responsible.
Fortunately, LLMs only occasionally made mistakes in this regard, as evidenced by one driver with 16 such modified identifier values, yet only one of them was inferred incorrectly by the LLM.
Lastly, only 9 syscalls out of 7 drivers exhibited incorrect types.
Overall, these results show that \tech can infer specifications with high accuracy and completeness.

\subsubsection{Bug Detection by New Specifications}

Table~\ref{tab:bug} shows the unknown kernel vulnerabilities detected using the newly generated specifications by \tech.
\tech has detected \numNewBug previously unknown bugs, with \numConfirmed confirmed by the kernel developers.
\numCVE of them are assigned with CVE numbers, and \numFix are already fixed.
Notably, \emph{none of them} can be detected by the default \syzkaller
or \syzdescribe since they are only triggered by the new descriptions generated by \tech, emphasizing the effectiveness of \tech in revealing real-world kernel bugs.
\numBugNewDriver bugs are detected from the drivers/sockets that
have been loaded in the default \CodeIn{syzbot} configuration for an extended period but \syzkaller lacks specifications for them.
Interestingly, the other \numBugNewSpec bugs were not revealed by \syzkaller because their specifications are incomplete.
For example, \syzkaller's descriptions for the RDS socket~\cite{rds} cover only the \CodeIn{recvmsg} syscall, omitting \CodeIn{sendto}.
By generating the missing \CodeIn{sendto} specification, \tech uncovered an array index out-of-bounds vulnerability in it,
which was acknowledged with a CVE and patched by kernel developers.
Next, we analyze and discuss two additional CVEs that are in the non-described drivers.

\parabf{CVE-2024-23848.}
This vulnerability, titled \textit{KASAN: slab-use-after-free Read in cec\_queue\_msg\_fh}, is found within the CEC driver, for which \syzkaller lacks descriptions.
It accesses a variable after it has been deallocated by \CodeIn{kfree(fh)}.
The issue stems from the driver's failure to properly maintain a lock while releasing resources, leading to a Use-After-Free vulnerability.
This bug has been rectified by kernel developers and has been assigned a CVE due to its exploitability.

\parabf{CVE-2024-23851.}
This bug, \textit{kmalloc bug in ctl\_ioctl}, is detected by \tech in the device mapper driver, which is also not described by \syzkaller.
The root cause is that the driver neglects to check the allocation size for \CodeIn{kvmalloc}, leading to the possibility of allocating excessively large memory sizes.
Specifically, the issue is associated with the \CodeIn{date\_size} field in the \CodeIn{dm\_ioctl} struct.
This field plays a crucial role in allocating memory during the preparation of the data structure within \CodeIn{copy\_param}.
These elements are key in the process of allocating targets while executing \CodeIn{dm\_table\_create}.
Notably, although \syzdescribe generates a specification for this driver, it incorporates an incorrect device filename, an erroneous command value, and imprecise types, thereby failing to detect this vulnerability.
\emph{Linus Torvalds confirmed this bug~\cite{linus_email} in addition to providing detailed fixing suggestions since it required an in-depth understanding of the entire Linux codebase}.
Given its potential for exploitation in DoS attacks, it has also been assigned a CVE.

\subsection{Specification Generation for Existing}

To further evaluate the quality of \tech-generated specifications in terms of fuzzing, we apply it to generate specifications for the ``existing'' drivers and sockets described by our baselines, \syzkaller and \syzdescribe~\cite{syzdescribe}, the state-of-the-art specification generation techniques.
We opt for all the 30 drivers used in the evaluation setting of \syzdescribe, as detailed in Table 6 of their paper~\cite{syzdescribe}.
For sockets, we compare against \syzkaller only.
Regarding \syzdescribe, it cannot analyze and generate descriptions for sockets,
attributed to the extensive implementation efforts required.
We randomly selected 10 socket handlers using a seed value of 0, after arranging them in alphabetical order.
We run each generated specification independently for 6 hours (48 CPU hours) with 3 repetitions to compare the coverage results.
During these runs, we specifically enabled only the syscalls included in the specification for each driver or socket.

\definecolor{mycolor}{rgb}{0.9,0.9,0.9}
\begin{table}[t]\centering
\caption{Comparison of driver specification generation with state-of-the-art solutions. \emph{\#Sys} is the number of syscalls described for the drivers. \emph{Cov} represents the average coverage.}
\label{tab:baseline}
\footnotesize
\begin{tabular}{l|rc|rc|rcr}\toprule
&\multicolumn{2}{c|}{\syzkaller} &\multicolumn{2}{c|}{\syzdescribe} &\multicolumn{2}{c}{\tech} \\\cmidrule{2-7}
&\#Sys &Cov &\#Sys &Cov &\#Sys &Cov \\\midrule
ashmem &N/A &- &N/A &- &N/A &- \\
btrfs-control &1 &1523 &5 &\cellcolor{mycolor}\textbf{2848} &5 &2786 \\
capi20 &13 &2818 &19 &3011 &14 &\cellcolor{mycolor}\textbf{3138} \\
controlC\# &22 &4666 &Err &- &15 &\cellcolor{mycolor}\textbf{4703} \\
fd\# &N/A &- &N/A &- &N/A &- \\
fuse &2 &1719 &2 &2315 &2 &\cellcolor{mycolor}\textbf{2425} \\
hpet &1 &1591 &7 &2289 &7 &\cellcolor{mycolor}\textbf{2493} \\
i2c-\# &10 &4168 &10 &4024 &10 &\cellcolor{mycolor}\textbf{4475} \\
kvm &118 &10948 &165 &9444 &71 &\cellcolor{mycolor}\textbf{15605} \\
loop-control &4 &7042 &4 &8211 &4 &\cellcolor{mycolor}\textbf{8537} \\
loop\# &12 &8498 &12 &\cellcolor{mycolor}\textbf{8519} &12 &8518 \\
mISDNtimer &3 &\cellcolor{mycolor}\textbf{1992} &3 &1965 &3 &1960 \\
nbd\# &11 &4103 &13 &5311 &12 &\cellcolor{mycolor}\textbf{5475} \\
nvram &1 &1618 &3 &2329 &6 &\cellcolor{mycolor}\textbf{2341} \\
ppp &24 &5710 &41 &6102 &34 &\cellcolor{mycolor}\textbf{7509} \\
ptmx &49 &\cellcolor{mycolor}\textbf{11598} &41 &10870 &30 &11344 \\
qat\_adf\_ctl &6 &2788 &6 &2651 &6 &\cellcolor{mycolor}\textbf{2883} \\
rfkill &3 &2117 &4 &\cellcolor{mycolor}\textbf{2388} &3 &2301 \\
rtc\# &24 &4458 &33 &4596 &17 &\cellcolor{mycolor}\textbf{5513} \\
sg\# &39 &\cellcolor{mycolor}\textbf{7412} &30 &6414 &43 &7392 \\
snapshot &13 &3076 &16 &3260 &15 &\cellcolor{mycolor}\textbf{3470} \\
sr\# &1 &2882 &68 &3725 &58 &\cellcolor{mycolor}\textbf{5091} \\
timer &16 &3328 &Err &- &17 &\cellcolor{mycolor}\textbf{3621} \\
udmabuf &4 &2771 &25 &2115 &4 &\cellcolor{mycolor}\textbf{2921} \\
uinput &22 &5470 &24 &4714 &21 &\cellcolor{mycolor}\textbf{6397} \\
usbmon\# &9 &3646 &16 &3806 &9 &\cellcolor{mycolor}\textbf{4332} \\
vhost-net &34 &\cellcolor{mycolor}\textbf{3615} &25 &3435 &22 &3541 \\
vhost-vsock &3 &2911 &25 &3448 &22 &\cellcolor{mycolor}\textbf{3803} \\
vmci &18 &3760 &26 &4316 &18 &\cellcolor{mycolor}\textbf{4674} \\
vsock &1 &1541 &2 &\cellcolor{mycolor}\textbf{1821} &2 &1744 \\ \midrule
Total &464 &117769 &625$^*$ &113927 &482 &\cellcolor{mycolor}\textbf{138992} \\
\bottomrule
\end{tabular}

\end{table}

\subsubsection{Device Drivers}

Table~\ref{tab:baseline} presents the results for drivers.
Due to space constraints, we omit the average number of unique crashes for each driver in Table~\ref{tab:baseline}. In total, \tech triggers 24.0 unique crashes, compared to 21.0 by \syzkaller and 20.7 by \syzdescribe.
Two baseline drivers, \CodeIn{ashmem} and \CodeIn{fd\#}, are no longer supported in Linux 6 (``N/A'').
Notably, \tech achieves the \emph{highest} \revision{basic block} coverage and crashes, surpassing the baselines by at least 18.0\% and 17.6\%.
Moreover, \tech performs the best on 20 of 28 (excluding the 2 invalid drivers, highlighted in bold), whereas \syzkaller and \syzdescribe lead in only 4 and 4, respectively.
This highlights the effectiveness of \tech-generated specifications in enhancing fuzzing.
For \CodeIn{kvm} driver~\cite{kvm}, \tech identifies two additional operation handlers, \CodeIn{kvm\_vm\_fops} and \CodeIn{kvm\_vcpu\_fops}, as dependencies. This leads to a coverage increase of \emph{42.5\%} and \emph{65.2\%} compared to baselines.

While \syzdescribe has the largest set of specifications, it repeatedly describes the same \CodeIn{ioctl} syscall using different types, which is atypical.
An \CodeIn{ioctl} command accepts only a single type in most scenarios.
Excluding duplicates, \tech describes more distinct syscalls (482) than \syzdescribe (464). Additionally, \syzdescribe incorrectly inferred device names for \CodeIn{controlC\#} and \CodeIn{timer}, preventing coverage.

\begin{table}[t]\centering
\caption{Comparison on socket specification generation.}
\label{tab:socket}
\footnotesize
\begin{tabular}{l|rcc|rccr}\toprule
&\multicolumn{3}{c|}{\syzkaller} &\multicolumn{3}{c}{\tech} \\\cmidrule{2-7}
&\# Sys. &Cov &Crash &\# Sys. &Cov &Crash \\\midrule
caif\_stream &4 &8947 &0.7 &6 &\cellcolor{mycolor}\textbf{11902} &0.7 \\
l2tp\_ip6 &38 &\cellcolor{mycolor}\textbf{18350} &0.7 &99 &18080 &0.7 \\
llc\_ui &10 &7648 &0.3 &24 &\cellcolor{mycolor}\textbf{16437} &0.0 \\
mptcp &22 &10480 &1.3 &70 &\cellcolor{mycolor}\textbf{13942} &0.7 \\
packet &22 &\cellcolor{mycolor}\textbf{22082} &0.3 &25 &21363 &0.3 \\
phonet\_dgram &7 &11426 &1.0 &12 &\cellcolor{mycolor}\textbf{15202} &0.7 \\
pppol2tp &10 &\cellcolor{mycolor}\textbf{18789} &0.3 &14 &12379 &0.7 \\
rds &11 &13693 &0.3 &19 &\cellcolor{mycolor}\textbf{17462} &1.0 \\
rfcomm\_sock &22 &7263 &1.0 &16 &\cellcolor{mycolor}\textbf{10893} &0.7 \\
sco\_sock &20 &11349 &1.0 &19 &\cellcolor{mycolor}\textbf{16527} &0.7 \\ \midrule
Total &166 &130027 &7.0 &304 &\cellcolor{mycolor}\textbf{154187} &6.0 \\
\bottomrule
\end{tabular}
\end{table}
\vspace{-5pt}

\subsubsection{Sockets}
Table~\ref{tab:socket} presents the results for sockets.
We observe that \tech can cover \emph{\moreSocketLine\%} more \revision{basic blocks} than our baseline \syzkaller, demonstrating the effectiveness of our generated specifications for sockets.
Notably, \tech describes significantly more syscalls than \syzkaller, showcasing \tech's capability in syscall discovery.
This is also because \syzkaller often uses a single syscall with various command values, while \tech generates distinct syscalls for each command value.
For example, \tech synthesizes 99 descriptions for the \CodeIn{l2tp\_ip6} socket handler, compared to \syzkaller's 38.
This is because the single \syzkaller syscall of this socket, \CodeIn{getsockopt\$inet6\_int}, uses \CodeIn{flags[inet6\_option\_types\_int]} as the command value \emph{list}, encompassing 45 unique syscall identifier values.

\subsubsection{Ablation Study}

We perform a comprehensive ablation study to evaluate how \tech's behavior would be affected if different components of it were disabled or modified.
Due to resource limitations, we selected only the first 10 \emph{valid} drivers from Table~\ref{tab:baseline}.

\parabf{Iterative Multi-Stage Generation.}
We experimented with a new setup where all function code related to syscalls was combined into a single prompt and the LLM generated the specification in one step, deviating from our iterative multi-stage generation process. This simplified approach, however, resulted in a noticeable decline in the quality of the generated specifications, especially for complex drivers like \CodeIn{kvm} and \CodeIn{loop\#}.
For example, iterative multi-stage prompting infers 71 syscalls and 28 types for \CodeIn{kvm}, while an all-in-one prompting approach infers only 42 syscalls and 11 types, resulting in a substantial decline in coverage (15,605 versus 5,457).
Overall, iterative multi-stage prompting can infer 1.28X more syscalls and 2.37X more types, leading to a 1.39X improvement in coverage for these 10 drivers.

\parabf{LLM Choice.}
We examined the performance of \tech when utilizing other \llm{s}, specifically \gpt{-3.5} and \gpt{-4o}. Our observations revealed that employing \gpt{-3.5} significantly reduced the number of described syscalls (85 versus 143), leading to a 21\% decrease in coverage compared to our default choice, \gpt{-4}.
Conversely, when using \gpt{-4o}, \tech was able to infer a similar number of syscalls (144 versus 143), and the coverage results were comparable to our default choice, \gpt{-4} (55,771 versus 54,640).
We would conclude that the syscall specification inference task necessitates a sufficiently powerful model like \gpt{-4} and \gpt{-4o}.

\section{Related Work}
\subsection{Kernel Fuzzing}
SyzGen~\cite{syzgen} infers syscall specifications but targets binary-only macOS drivers, leveraging symbolic execution to recover the data types and syscall traces to find dependencies.
Moonshine~\cite{moonshine} collects and distills syscall traces to generate a seed pool for \syzkaller.
SyzVegas~\cite{syzvegas} leverages reinforcement learning to dynamically improve seed and task selection.
HEALER~\cite{healer} infers syscall dependencies by observing coverage changes with different combinations.
Plus, SyzDirect~\cite{syzdirect} applies directed grey-box fuzzing for \syzkaller by incorporating distance information.
PrIntFuzz~\cite{printfuzz} uses static analysis to extract driver information and generate their simulators for fuzzing.
Focusing on synthesizing specifications, \tech is orthogonal to the above techniques and could be combined to improve \syzkaller collectively.

\subsection{\revision{Learning-Based Fuzzing}}

\revision{Machine learning approaches to input generation~\cite{godefroid2017learn,liu2019deepfuzz,Rajpal2017NotAB,Sablotny2018RecurrentNN,cummins2018compiler} explored using sequence-to-sequence models to learn program syntax and generation patterns. For instance, Learn\&Fuzz~\cite{godefroid2017learn} leveraged sequence-to-sequence models for grammar-based fuzzing by learning from sample inputs. DeepFuzz~\cite{liu2019deepfuzz} extended this approach to generate C programs for compiler testing. DeepSmith~\cite{cummins2018compiler} also demonstrated the potential of learning program patterns from a large corpus. However, these approaches faced significant limitations: they required extensive domain-specific training data, exhibited lower throughput compared to traditional fuzzers, and struggled to leverage existing fuzzing infrastructure. Our initial explorations show that even using more powerful \llm{s} for direct input generation in kernel fuzzing performed notably worse than basic \syzkaller.}

Recent advancements have shown that \llm{s}~\cite{chatgpt,gpt4,starcoder,codex,codellama} excel in a variety of natural language processing~\cite{brown2020language} and programming tasks~\cite{xu2022systematic,bubeck2023sparks, xia2024agentless, xia2023automated}.
Their proficiency in diverse tasks is attributed to the extensive training on vast datasets, \eg{} \gpt{4}~\cite{gpt4} is pre-trained using trillions of text tokens from the entire Internet.
As a result, \llm{s} can be employed in various tasks simply by following instructions~\cite{ouyang2022training,bang2023multitask,chatgpt}, eliminating the need for specialized training.
Recently, a growing body of research has focused on leveraging \llm{s} for software testing, covering both unit test generation~\cite{nie2023learning,lemieux2023codamosa,schafer2023adaptive,yuan2023no} and fuzzing~\cite{fuzz4all,titanfuzz,yang2023white,hu2023augmenting,chatafl}.
For example, \titanfuzz~\cite{titanfuzz} pioneered the application of modern \llm{s} for both generation-based~\cite{csmith,yarpgen} and mutation-based~\cite{donaldson2017automated,le2014compiler} fuzzing, while
\fuzzall~\cite{fuzz4all} further demonstrated that the multilingual potentials of \llm{s} can be utilized to serve as a universal fuzzer for a wide range of software systems.

\revision{\tech takes a fundamentally different approach from both traditional ML-based and recent \llm-based fuzzing techniques. Instead of directly generating test inputs, we integrate \llm{s} with mature fuzzing frameworks by synthesizing their \emph{components} (\ie input generators). This strategy leverages the expertise and resources invested in well-developed fuzzing tools while capitalizing on \llm{s}' capabilities. To the best of our knowledge, \tech is the first work to successfully apply \llm{s} to kernel fuzzing, demonstrating state-of-the-art performance in generating high-quality syscall specifications.}

\section{Conclusion}
In this paper, we propose \tech, the first approach to synthesizing syscall specifications automatically via \llm{s} for enhanced kernel fuzzing.
It employs an iterative method to autonomously deduce syscall specifications and further repair them using validation feedback.
Experimental results show that \tech helps improve \syzkaller's coverage and can detect \numNewBug previously unknown bugs through the newly generated specifications, with \numCVE CVE assignments and \numFix fixed.
Additionally, a number of specifications inferred by \tech are already merged into \syzkaller repository, following a request from its development team. 
To our knowledge, this is the first automated approach to leveraging \llm{s} for kernel fuzzing.
It could open up numerous possibilities for future research in this critical application domain.

\section*{Acknowledgments}

We are grateful to the anonymous reviewers and our shepherd, Youngjin Kwon, for their valuable feedback that helped improve this paper. This work was partially supported by NSF grant CCF-2131943 and Kwai Inc. Chenyuan Yang was partially supported by Boeing for research on Linux kernel testing. We also thank Ziqi Zhang for his helpful suggestions on the manuscript.

\vfill\eject
\bibliographystyle{plain}
\balance
\bibliography{main}

\end{document}